%% file: paper.tex
\documentclass[sigconf]{acmart}
\usepackage{amsmath}
\usepackage{xspace}
\usepackage{multicol}
\usepackage{graphicx}
\usepackage{multirow}
\usepackage{comment}
\usepackage{natbib}
\usepackage{subfig}
\usepackage{xcolor}
\usepackage{csquotes}
\setcopyright{iw3c2w3}
\usepackage{balance}

\newcommand{\pa}{\textbf{political ads}\xspace}
\newcommand{\spa}{\textbf{strong political ads}\xspace}
\newcommand{\pnpa}{\textbf{marginally political ads}\xspace}
\newcommand{\npa}{\textbf{non-political ads}\xspace}
\newcommand{\opa}{\textbf{official political ads}\xspace}
\newcommand{\onpa}{\textbf{official non-political ads}\xspace}

\newcommand{\fr}{\textit{fr}\xspace}

\usepackage{lipsum}
\usepackage{xcolor}
\usepackage{soul}
\sethlcolor{black}

\newcommand{\AdAnalyst}{AdAnalyst\xspace}

\begin{document}

\copyrightyear{2021}
\acmYear{2021} 

\acmConference[WWW '21]{Proceedings of the Web Conference 2021}{April 19--23, 2021}{Ljubljana, Slovenia} 
\acmBooktitle{Proceedings of the Web Conference 2021 (WWW '21), April 19--23, 2021, Ljubljana, Slovenia}
\acmPrice{}
\acmDOI{10.1145/3442381.3450049}
\acmISBN{978-1-4503-8312-7/21/04}

%\title{Political Ad Detection: Who should label the ads?}
\title{Understanding the Complexity of Detecting Political Ads}
%\title{Political Ad Detection: why is it hard?}

\author{Vera Sosnovik}
\affiliation{% 
  \institution{Univ. Grenoble Alpes, CNRS,
Grenoble INP, LIG}
\country{France}
}

\author{Oana Goga}
\affiliation{%
  \institution{Univ. Grenoble Alpes, CNRS,
Grenoble INP, LIG}
\country{France}
}

\begin{abstract}

Online political advertising has grown significantly over the last few years. To monitor online sponsored political discourse, companies such as Facebook, Google, and Twitter have created public Ad Libraries collecting the political ads that run on their platforms. Currently, both policymakers and platforms are debating further restrictions on political advertising to deter misuses.  

This paper investigates whether we can reliably distinguish political ads from non-political ads. We take an empirical approach to analyze what kind of ads are deemed political by ordinary people and what kind of ads lead to disagreement. %, and whether there are differences between what advertisers consider political and what ordinary people consider political. 
Our results show a significant disagreement between what ad platforms, ordinary people, and advertisers consider political and suggest that this disagreement mainly comes from diverging opinions on which ads address social issues. 
Overall our results imply that it is important to consider social issue ads as political, but they also complicate political advertising regulations.

% form social issue ads.  need for more narrow and precise definitions for ads about social issues in order to be able to 
%definitions for social issue we cannot reliably detect political ads. \update{Hence, applying restrictions on political ads when there is no clear cut between political and non-political ads would be problematic.} 

% This paper will show the importance of considering social issue ads as sensitive content and why they complicate political advertising regulations. 

\end{abstract}

\maketitle

\input{intro}
%\input{background}
\input{dataset}
\input{agreement}
\input{political_ads}
\input{sources_disagreement}
\input{classifiers}

%\input{prolific}
\input{relatedworks}
\input{conclusions}

\section{Acknowledgements}
We thank the anonymous reviewers for their helpful comments and Jeremy Merrill for his assistance with the data.
This research was supported in part by the French National Research Agency
(ANR) through the ANR-17-CE23-0014 and the MIAI@Grenoble
Alpes ANR-19-P3IA-0003 grants.

\balance

\bibliographystyle{ACM-Reference-Format}
\bibliography{biblio}

\input{appendix}

\end{document}

%% file: intro.tex
%!TEX root = paper.tex

\section{Introduction}

Social media and the public sphere's digitalization have changed the political campaigning landscape in both good and dangerous ways. While social media are creating new opportunities for engaging citizens in political conversations, they have also raised several risks for the integrity of elections and the political debate.  
For example, online ads can be tailored to specific groups of people, hence polarizing the voter base and distorting the political debate. Advertisers can buy large amounts of ads to flood people's social media feeds and steer public debates on issues that are of interest to them. Anyone, from political parties to interest groups, and specialized advertising companies such as Cambridge Analytica, can steer the political debate through online advertising. 

Ad platforms have put forward several measures to mitigate risks and allow for public scrutiny of ads. Twitter and TikTok decided to ban political ads altogether. Google and Facebook allow political ads, but advertisers are subject to a higher degree of scrutiny and limitations. On Google, advertisers can only use geographic location, age, gender, and contextual targeting to target political ads.  Facebook does not restrict the micro-targeting of political ads. Advertisers, however, need to verify their account (by showing proof of  identity or a public listing of their business~\cite{facebook-political-ads}) and are only allowed to send political ads to users that reside in the same country as them. Moreover, advertisers have to self-declare when their ads are political, and all political ads sent on the platform appear in the Facebook Ad Library where the civil society can further scrutinize them~\cite{facebook-ad-library}. 

On their side, governments are also working on legislation to regulate political advertising. 
For example, the European Commission is working on two pieces of legislation, the Digital Services Act (DSA)~\cite{DSA} and the European Democracy Action Plan (EDAP)~\cite{EDAP}, that aim to regulate in part online advertising. In a recent interview, V\v{e}ra Jourov\'a (the European Commission Vice-President for Values and Transparency),  declared~\cite{Jourova}:
 \begin{displayquote}
{\textit{... we are coming in the European Democracy Action Plan with the vision for the next year to come with the rules for political advertising, where \textbf{we are seriously considering to limit the microtargeting as a method which is used for the promotion of political powers, political parties or political individuals.}}}
\end{displayquote}
%\update{Other advocacy groups are militating to ban altogether political advertising or micro-targeting for all ads~\cite{XX}.}  
Measures from both ad platforms and governments are positive developments. However, all of them implicitly rely on the assumption that \textit{one can reliably distinguish political ads from non-political ads}. 

In this paper, we take an empirical approach to test this assumption by analyzing the characteristics of ads deemed political by ordinary people, the characteristics of ads that lead to disagreement, and whether there are differences between what advertisers consider political and what ordinary people consider political.
%to understand what are the ads that are considered political by actual people, and more importantly  
%ads are considered political by people, and more importantly, what ads lead to disagreement. 
Our analysis is based on a dataset from ProPublica that contains 55k Facebook ads received by U.S. residents, labeled by at least one volunteer as political, and that received three or more votes (Section~\ref{sec:dataset}). The dataset was collected by a browser extension that collects the ads users see when they browse their Facebook timeline and allows users to label whether the ads they see are political.  

First, we investigate whether ad platforms, volunteers, and advertisers agree on which ads should be considered political (Section~\ref{sec:agreement}). All ad platforms agree that ads from or about political actors and ads about elections and voting should be considered political. However, only Facebook and TikTok consider ads about \textit{social issues} (such as climate change or immigration) as political. Our results show that volunteers disagree on whether an ad is political for more than 50\% of the ads in the dataset, and only 83\% of the ads labeled as political by advertisers are also labeled as political by a majority of volunteers. Hence, the fundamental assumption that we can clearly distinguish political from non-political ads does not hold, since there is no consensus even on what constitutes a political ad, and volunteers and advertisers label different sets of ads as political. 

%there is significant disagreement between ad platforms, volunteers, and advertisers. %Therefore, applying restrictions on political ads when there is no clear cut between political and non-political ads is problematic. % because it can leads to unfairness between advertisers.

Next, we analyze the characteristics of ads that are labeled as political by volunteers and advertisers in the ProPublica dataset, which can be useful to inform the debate on definitions of political ads (Section~\ref{sec:political_ads}). To that end, we gathered data about the advertisers sending political ads and the content of their ads. We hired Prolific users to annotate 2300 ads with the political or social issues the ad is referring to. Our analysis shows that a wide range of advertisers (from political actors to NGOs and businesses) are posting political ads on Facebook and that ads about social issues account for a large fraction of the ads labeled as political; hence emphasizing the importance of including such ads in political ads definitions. 
Our analysis also shows that the ads labeled as political by volunteers and advertisers are very diverse. We see ads with a clear political message from advocacy groups (e.g., ads addressing abortion issues in the U.S.); but also ads from NGOs that address humanitarian issues and do not seem to directly or indirectly impact U.S. elections or legislation (e.g., ads asking for donations for ending world hunger). 
As political ads may be subject to higher restrictions, this questions whether it is desirable that the same restrictions apply to both types of ads. More generally, this emphasizes the need to account for the diversity of political ads in devising regulations. 
%Hence, since political ads will potentially be subject to higher restrictions, we need to decide whether we want ads (with no apparent link to elections and legislation) from charities or communities to be subject to the same restrictions as ads from advocacy and political groups. 
 
We finally analyze the ads that lead to disagreement among volunteers and between volunteers and advertisers (Section~\ref{sec:sources_disagreement}). 
We first observe that advertisers mislabel some ads as either political or non-political (according to the Facebook ToS). 
%Then, we find an important fraction of ads deemed political by volunteers that are unreported by advertisers for some social issues (in particular economy and civil and social rights), and vice-versa for certain social issues (in particular health and also civil and social rights) and for ads from advertisers such as NGOs and charities. 
Then we find that advertisers seem to underreport ads (that are considered political by volunteers) about social issues, especially the economy and civil and social rights. Volunteers seem to underreport ads (considered political by advertisers) from advertisers such as NGOs and charities, and about social issues, especially civil and social rights and health. 
Part of the problem may be that the definition of ads about social issues may be too broad and vague, which leads to being interpreted in different ways by people. This also raises the question of whether \textit{all} ads related to social issues should be considered political, and if not, how to filter social issue ads that are not political.

Because of the high volume of ads, enforcement mechanisms need to rely on automated machine learning (ML) algorithms to detect political ads. However, it is not clear how one should train and evaluate such models since there is disagreement on which ads are political (i.e., the positive examples).
To investigate that, we train four classifiers with different groups of positive examples (coming from advertisers and volunteers). We test how they perform over various groups of political ads with varying degrees of disagreement (Section~\ref{sec:classifier}). While all classifiers achieve high accuracy in detecting ads everyone agrees are political; their accuracy drops on ads that only a few find political. 

Another important question is whether (and to which extent) models trained with labels from advertisers would declare as political the same ads as models trained with labels from volunteers (i.e., reliable detection of political ads). Theoretically, if ads labeled as political by advertisers and volunteers are representative of political ads in general, the resulting models should declare the same ads as political. Our results show that the overlap between different models is relatively high (ranging from 82\% to 97\%), but that discrepancies in the input data transfer to discrepancies in the output data. This suggests that  existing labeled datasets are not providing a representative set of political ads needed to build reliable detection schemes. 

% However, since there is disagreement on what ads should be considered political, it is not clear how we should assess the accuracy of the resulting models, and how we should account for disagreement in the input data.
%Section~\ref{sec:classifier} investigates the accuracy of four ML-models, one trained with ads labeled as political by advertisers, and three trained with ads labeled as political by volunteers with varying degrees of disagreement. 

%how can we account for this confusions and how can we assess accuracy of the resulting models. 
%The accuracy of these models depends on the quality and quantity of training data. \textit{The question is when dealing with political ads since it is not always clear what ads are political, how to assess the quality of the labels and the accuracy of models.} 

%Hypothetically if the resulting classifiers end up labeling the same ads as political, then who labeled the training data does not matter. Our results show that both the advertiser-based classifier and the volunteer-based classifier labeled as political 31k ads, and 2k ads were only labeled by one classifier as political.  

Overall, our work suggests that, given the complexity of deciding which ads are political, it would be beneficial to have ad libraries that contain \textit{all} ads running on the platform, not only ads deemed political by the ad platform. Following this work, we issued a statement together with civil societies asking for ``\textit{Universal advertising transparency by default}'' that we submitted to the DSA consultation~\cite{UVT}. However, this crucial first step is not enough because political ads are also subject to higher restrictions; hence, we still need to detect political ads reliably. We hope this study can help policymakers to define political speech and decide on appropriate restrictions and ad platforms to set infrastructures for detecting political ads.

%We believe that, given the complexity of deciding what ads are political, it is essential to have ad libraries that contain \textit{all} ads running on the platform, not only ads deemed as political by the ad platform. Together with more than 30 civil societies, we wrote a statement asking for ``\textit{Universal advertising transparency by default}'' that we submitted to the DSA consultation~\cite{UVT}. The December 2020 DSA draft seems to go in this direction. 

%policymakers and ad platforms to devise adapted definitions and guidelines to alleviate disagreement and confusion in the ad labeling process. 

%% file: dataset.tex
%!TEX root = paper.tex

\section{Datasets}
\label{sec:dataset}
For our analysis we use the following two datasets of ads that users have received on their Facebook timeline: 

\begin{table}[t]
\small{
\caption{\label{tab:nbads} Number of ads in the ProPublica and the     AdAnalyst datasets, and percentage of ads with official ``Paid for by'' political disclaimer.}
\vspace{-2mm}
 \begin{tabular}{|l c c c|} 
 \hline
  & All ads & Official political & Official non-political \\
 \hline\hline
 ProPublica & 54.6k & 50.8k (93\%) & 3.8k (7\%)\\
\hline
AdAnalyst&9k&2\%&98\%\\
\hline
\end{tabular}
}
\vspace{-5mm}
\end{table}

\vspace{-2mm}
\paragraph{ProPublica dataset} 
ProPublica, an investigative journalism organization, has developed a browser extension that collects the ads users are receiving on Facebook and allows users to label whether the ads they are seeing are political or not~\cite{pb-data}. The extension is currently maintained by the NYU Online Political Transparency Project~\cite{NYU-pr}. While ProPublica was not able to make available all the ads it has collected, it shared with us \textbf{all the ads} \textbf{for which at least one user has labeled it as being political}, as well as \textbf{all the ads that have the ``Paid for by'' disclaimer} (i.e., the official political ads that have been declared as such by advertisers). This dataset is valuable because it provides us with a unique view of which ads are considered political by ``ordinary'' people/volunteers. To our knowledge, there are no studies of such data. 

%The dataset contains ads from September 2017 to May 2020.

For this study, we only kept ads with at least three votes (either political or non-political) and that were received  between June 2018 and May 2020; resulting in a dataset of 54.6k ads coming from 7530 advertisers. The median number of votes per ad after filtering is 5.  
We call the ads that have the ``Paid for by'' disclaimer the \opa and the ads that do not have the disclaimer the \onpa. Table~\ref{tab:nbads} shows the number of ads in the ProPublica dataset as well as the fraction of \opa and \onpa. Note that this dataset does not contain a representative sample of political ads as they are ads received by people who answered ProPublica's call for action to install the tool. 

%For our approaches, we used a dataset from ProPublica. This dataset contains more than 200 thousand ads that were automatically collect from Facebook by their extension. Moreover, this dataset has information about how many people labeled each ad as political and not political. We used this information for a volunteer-based approach. 

\vspace{-2mm}
\paragraph{\AdAnalyst dataset}
Similar to the extension provided by ProPublica, \AdAnalyst collects the ads users see on their Facebook timeline~\cite{andreou2019measuring}. The \AdAnalyst dataset contains over 500k ads from users in various countries. For this study, we keep only ads in English (detected using text-blob python library~\cite{text-blob}) and that targeted users in the US between October 2018 and May 2020. For this, we use information about ad targeting available in the ``Why am I seeing this ad?'' button and select only ads targeted at people who live in the USA or visited places in the USA recently.
The resulting dataset contains 9k unique ads (198 ads with ``Paid for by'' disclaimer and 8802 without). This dataset does not have votes from volunteers. 

\vspace{-2mm}
\paragraph{Ethical review board and reproducibility}
Both data collection by ProPublica and \AdAnalyst were approved by the respective ethical review boards. The ProPublica data is available to the public through a request form~\cite{pb_dataset}. The 9k ads from \AdAnalyst, the data collected from the Prolific studies, and other supplemental material is available at \url{http://lig-membres.imag.fr/gogao/www21.html}.

%We will make public the 9k ads from \AdAnalyst used in our study and the data collected from our Prolific studies. All results in the paper will be reproducible. \oana{add link to website where data will be available}

%% file: agreement.tex
%!TEX root = paper.tex

\section{Disagreement on political ads}
\label{sec:agreement}
The base to detect political ads reliably is to agree on which ads should be considered political and which ads should not. In this section we look at whether ad platforms, volunteers, and advertisers agree on which ads are political. 

% In this section, we will investigate whether ad platforms agree among them on what ads should be treated as political, whether volunteers agree amongst them what ads are political and if there are differences in what ads are labeled as political by advertisers and volunteers. 

\subsection{Disagreement across ad platforms}
%\oana{Do platforms agree on what ads are political?} 

The Terms of Services of different ad platforms provide information on which ads they consider as political. We review the definitions and restrictions for political advertising across four ad platforms. 

\vspace{1mm}
\noindent \textbf{Facebook} defines political ads as: \textit{
``Made by, on behalf of, or about a candidate for public office, a political figure, a political party, or advocates for the outcome of an election to public office; 
 About any election, referendum, or ballot initiative, including "go out and vote" or election campaigns; 
 About social issues in any place where the ad is being placed; 
Regulated as political advertising.''}
The social issues are: \textit{civil and social rights, crime, economy, education, environmental politics, guns, health, immigration, political values and government, security, and foreign policy}~\cite{facebook-political-ads}.

Everyone with a Facebook account can be an advertiser if they provide a payment method. However, to be able to send political ads, advertisers need to verify their accounts by providing proof of their identity~\cite{facebook_adv}. Advertisers can only send political ads in the country they reside and need to provide proof of residence. Advertisers need to self-label their ads as political and need to provide a 
disclaimer about who paid for the ad. This ``Paid for by'' disclaimer appears on the top of the ad frame, after the advertiser's name. Finally, Facebook adds the political ads to their Ad Library~\cite{facebook-ad-library}.

\vspace{1mm}
\noindent \textbf{Google} defines political ads as: \textit{``ads about political organizations, political parties, political issue advocacy or fundraising, and individual candidates and politicians''}~\cite{google-political-ads}. 
The platform imposes no restrictions on political ads, but the platform expects all political ads to comply with local legal requirements. 
Google considers \textit{election ads} as a separate category. The definition of election ads depends on the country, but overall it refers only to ads from or about candidates and political parties during an electoral period. Only verified advertisers can run election ads. Election ads can only be targeted by geographic regions (but not by radius around a precise location), age, gender, and contextual targeting options such as ad placements, topics, keywords against sites, apps, pages, and videos. 

\vspace{1mm}
\noindent \textbf{Twitter} defines political ads as \textit{``ads with political content: that references a candidate, political party, elected or appointed government official, election, referendum, ballot measure, legislation, regulation, directive, or judicial outcome; as well as ads of any type by candidates, political parties, or elected or appointed government officials''}~\cite{twitter-political-ads}. 
Twitter bans all political ads.  
 
\vspace{1mm}
\noindent \textbf{TikTok}
defines political ads as \textit{ads that promote or oppose a candidate, current leader, political party or group, or issue at the federal, state, or local level — including election-related ads, advocacy ads, or issue ads}~\cite{TikTok}. 
TikTok bans all political ads.  

Overall there are three categories of political ads: \textbf{ads from or about a political figure or political party}, \textbf{ads about elections}, and \textbf{ads about social issues}. 
While the precise definition of political ads varies across ad platforms, the most significant difference is that Twitter and Google do not consider ads about social issues as political while Facebook and TikTok do. While it is certainly a debatable question whether or not social issue ads should be regarded as political, the EU Code of Practice on Disinformation mentions both issue ads and political ads as sensitive content. Our results will show the importance of considering social issue ads as political and why they complicate political advertising regulations. 

 %\update{The paper will bring evidence of why we should consider social issue ads as political and what are the drawbacks.}

%\oana{While small differences between definitions might not seem very problematic, it is problematic when we apply restrictions to political ads; such that no micro-targeting}

%On the legislative side, countries have specific election legislation that regulates political advertising before elections. These laws usually only refer to ads from political candidates and ads only around elections. 

\subsection{Disagreement among volunteers}
At least three volunteers have labeled each ad in the ProPublica dataset as being political or non-political. 
The volunteers were given no instructions for what ads they should consider as political, and users were left to decide based on their instinct. %This provides us with a perspective on what ads people intrinsically view as political, and that can help craft better definitions and instructions for what is a political ad. 

\begin{figure}[t]
 \centering
 \includegraphics[scale=0.4]{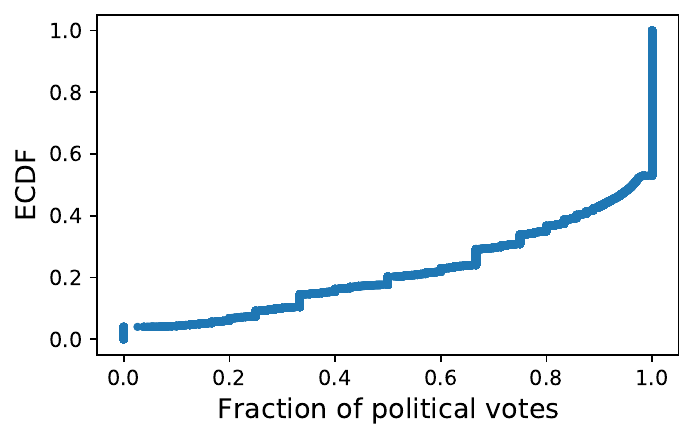}
 \vspace{-2mm}
 \caption{ECDF of the fraction of political votes for the ads in the ProPublica dataset.}
  \vspace{-3mm}
 \label{fig:fr_votes_new}
\end{figure}

%To understand the ads on which people agree and the ads on which people disagree, we look at \textbf{consensus} between volunteers. We consider agreement is a proxy for how complex the task of labeling political ads is. 
To observe to which extent volunteers agree on what ads are political, Figure~\ref{fig:fr_votes_new} plots the distribution of the number of political votes divided by the number of all votes for each ad in the ProPublica dataset. We denote this fraction as \fr. A fraction \fr$=1$ means that everyone agrees that the ad is political, while a fraction \fr$=0$ means that everyone agrees that the ad is not political. %
The plot shows that for more than 50\% of the ads, at least one volunteer disagrees with the others (\fr is neither 0 nor 1), which shows that \textit{deciding whether or not an ad is political is debatable for more than half of the cases.} 

To distinguish ads on which users agree they are political from the rest, we split the ads into four disjoint \textit{ad groups} based on the volunteer votes. We will analyze them separately in the paper. The groups are defined as follows:
\vspace{-1mm}
\begin{itemize}
\item \textbf{\spa}: ads with  $\textit{fr}=1$, i.e., where everyone agrees that they are political; 
\item \textbf{\pa}: ads with$0.5\leq \textit{fr} <1$, i.e., where there is some disagreement, but the majority labels them as political;
\item \textbf{\pnpa}: ads with  $0< \textit{fr} <0.5$, i.e., where there is some disagreement, but the majority labels them as non-political;
\item \textbf{\npa}: ads with $\textit{fr}=0$, i.e., where everyone agrees that are non-political.
\end{itemize}
%\vspace{-1mm}
There are 26k \spa, 19.7k \pa, 7.6k \pnpa, and 1.3k \npa.

 \begin{table}[t]

 \small{
 \caption{\label{tab:overlap} Number of ads in different ad groups (based on volunteer votes) and overlap with ads labeled as political by advertisers. $\dagger$ ProPublica was not able to give us access to ads that did not have at least one political vote and that were not labeled as official political ads.} 
 \vspace{-2mm}
 \begin{tabular}{|l c c c|} 
 \hline
 & All & Official pol. & Official non-pol. \\ %[0.35ex] 
\hline\hline
\spa & 26k &96\%& 4\%\\
 \hline
\pa & 19.7k& 93\%&7\%\\
 \hline
\pnpa & 7.6k & 74\% & 26\%\\
 \hline
\npa & 1.3k & 100\% & NA$\dagger$\\
\hline
\end{tabular}
}
\vspace{-4mm}
\end{table}

\subsection{Disagreement between volunteers and advertisers}
%\label{sec:labels advertiser}

The ProPublica dataset provides data on whether an ad was labeled as political by the advertiser itself (see Section~\ref{sec:dataset}). %Hence, we compare ads marked as political by advertisers and ads labeled as political by volunteers. 
Table~\ref{tab:overlap} presents the overlap between ads labeled as political by volunteers and ads labeled as political by advertisers (the \opa). The table shows that 96\% of \spa, and 93\% of \pa were also declared as political by advertisers. Hence, most ads considered political by the majority of volunteers are also considered political by advertisers. There are, however, 4\% of \spa and 7\% of  \pa that advertisers did not label as political. 

The more surprising finding is that advertisers label as political a large majority (74\%) of \pnpa. Looking the other way around, 83\% of \opa are labeled as political by most volunteers. In contrast, 15\% of \opa are only labeled as political by a minority of volunteers, and 2\% of \opa are not labeled as political by \textit{any} volunteer. Hence, many ads considered political by advertisers are not regarded as political by volunteers.  
While the reasons can be diverse (this is the subject of Section~\ref{sec:sources_disagreement}), we conclude that \textit{there is currently a significant discrepancy between the ads labeled as political by advertisers and by volunteers.}  

\vspace{1mm}
 \textbf{Takeaway:} The assumption that we can clearly distinguish political from non-political ads does currently not hold as there are significant disagreements between ad platforms, volunteers, and advertisers on which ads are political. Therefore, it is problematic to apply restrictions on political ads if the decision of whether an ad is political depends on the person labeling it.

%and what ads get labeled as political depends on who labeled the ads. 

%This is a bit surprising as we expected more extensive overlaps for \spa than \pnpa (i.e., what is considered highly political by volunteers will also be considered highly political by advertisers). Even more confusing, there are 1.3k ads where volunteers have consensus that they are non-political, but advertisers label them as political. Hence, \textit{there is a significant discrepancy between what advertisers labeled as political and what volunteers labeled as political.} {\color{red}{No longer relevant}}

%Table~\ref{tab:overlap} presents the number of ads in each group. 

%% file: political_ads.tex
%!TEX root = paper.tex

\begin{figure*}[ht]
 \centering
 \includegraphics[scale=0.4]{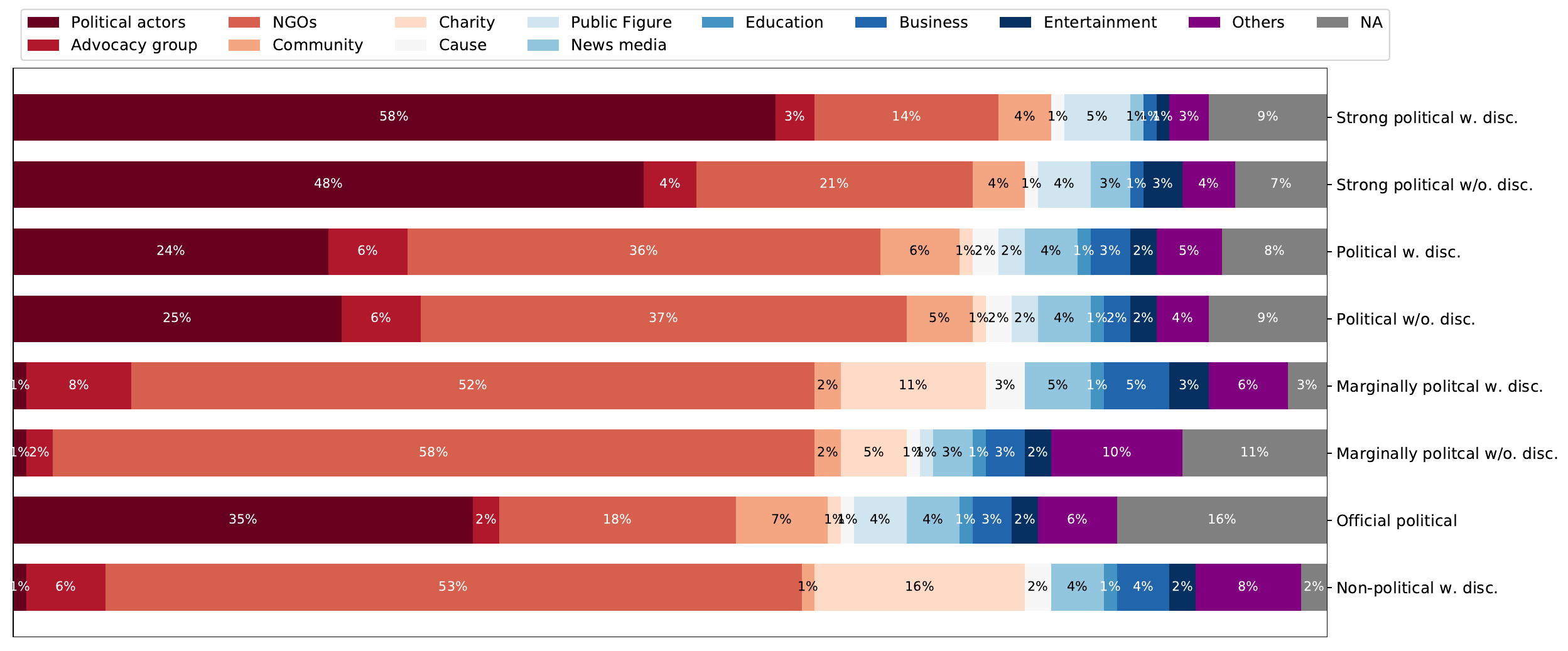}  
 \vspace{-4mm}
 \caption{Breakdown of advertisers categories for different groups of ads for ads with and without ``Paid for by'' disclaimer.}
 \label{fig:all_categories_disclaimer_new}
% \vspace{-2mm}
\end{figure*}

\section{What gets labeled as political}
\label{sec:political_ads}
This section provides a general view of ads labeled as political by volunteers and advertisers and analyzes who sends them and what are they talking about. This analysis is relevant for informing the debate on definitions of political ads and understanding the impact of potential regulations.
The next section will focus on which ads lead to disagreement. 

To interpret the results, we need to know the precise conditions in which the labeling happened.
The ProPublica \textit{volunteers} were given no instructions for what ads they should consider as political, and they were left to decide based on their subjective beliefs and background knowledge. However, volunteers could see if an ad was labeled as political by the advertiser itself (these ads have a ``Paid for by'' disclaimer on Facebook). We present results separately for ads that run with a disclaimer and ads that run without a disclaimer to isolate the potential effect of the ``Paid for by'' disclaimer. %Hence, we will see what ads people \textit{intrinsically} consider political. % which could help craft better definitions and instructions for labeling political ads

\textit{Advertisers} have to self-declare if they send political ads (as defined by Facebook's ToS). However, there is no public information on how Facebook enforces this policy~\cite{silva2020facebook}. 
Hence, ads labeled as political by advertisers are either a product of their own belief that their ad is political; or the result that Facebook constrained them to label the ad as political to run on the platform (maybe due to false positives in their enforcement algorithms).

\begin{table*}
\caption{Examples of ads from advertisers with different categories.}
\vspace{-2mm}
\scriptsize{

 \begin{tabular}{|p{0.15\textwidth}|p{0.75\textwidth}|p{0.02\textwidth}|p{0.015\textwidth}|}
 \hline
 Advertiser&Text & \fr & disc. \\  
 \hline
 
 \multicolumn{3}{c}{\textbf{Category: Cause}} \\
 \hline
 UnRestrict Minnesota& 96\% of Minnesotans don't know the abortion laws in our state. & 1 &w.\\
 \hline
 Care2& U.S. Wildlife Services is putting the safety of people: animals at risk in its attempt to control wild predators. Tell them to STOP using taxpayers money to kill wild animals lethally!& 0.75 & w.\\
 \hline
 Oregon Forests Forever & Brave men and women from Oregon are helping to fight fires in California& 0.37 & w.\\
 \hline
 Home Ownership Matters& Do you want Congress to invest in infrastructure? Click here to sign the petition.&0.33 &w.\\
 \hline
 
 \multicolumn{3}{c}{\textbf{Category: Charity}} \\
 \hline
 USA for UNHCR&Should America turn away from this child? Not now, not ever. It's not who we are.& 1& w.\\
 \hline
World Food Programme&I call on warring parties to allow the constant flow of food for innocent and starving people in Yemen. Add your voice to our petition today.&0.66 & w/o. \\
 \hline
ChildFund International & She wants a childhood free of worry and a future full of promise.& 0 & w.\\
 \hline
 USA for UNHCR& All donations MATCHED for a limited time. People in Syria are still fleeing for their lives. UNHCR needs your help to provide the shelter, food and medicine they need to survive.& 0.33 & w/o.\\
 \hline

 \multicolumn{3}{c}{\textbf{Category: Community}} \\
 \hline
Yes for Washington Elementary Students&Vote YES on the WESD Override to protect full-day kindergarten, music, art, and physical education in our schools.&1 & w.\\
 \hline
 North Carolina Citizens & We have a new survey for North Carolina. Please click the link below to share your thoughts& 0.8&w/o.\\
 \hline
 Healthy Me PA&Workplace violence is 4x more common in the health care industry. Here's how you can help:& 0.3& w. \\
 \hline
 Protect Coyote Valley&Time and time again, threats of development have been made in Coyote Valley, with some succeeding. We want to see Coyote Valley permanently protected for our wildlife and for our children. All we need is your signature& 0.4 & w. \\

 \hline
  \multicolumn{3}{c}{\textbf{Category: Business}} \\
 \hline
Dissent Pins&Stand for democracy on election day and every day with our Count Every Vote pin.&1 & w.\\
 \hline
 Ben and Jerry's & Vote YES on 4 and reinstate voting rights for 1.4 million Floridians!& 0.96&w.\\
 \hline
 CREDO Mobile&Help us decide how to allocate our \$50k donation to 5 progressive environmental organizations fighting for climate justice.& 0.33& w. \\
 \hline
 Steady Returns, LLC&Everyone deserves great financial advice!&0.2 & w/o.\\
 \hline
  
  \multicolumn{3}{c}{\textbf{Category: NGOs}} \\
 \hline
Democratic Attorneys General Association &Now that we know Joe Biden will be the nominee, we want to know who you think he should pick as his V.P.? Hurry, this round closes soon and we are still missing your response.&1 & w.\\
 \hline
 Pennsylvania Spotlight & Voting from home is easy. By taking thirty seconds to request a ballot, you can fill your ballot out on your couch and mail it in.& 0.63&w.\\
 \hline
 National Audubon Society&Birds and their habitats are under attack, but with your help we can fight back. This Earth Day your monthly gift will go twice as far to protect birds and the places they need..& 0.25& w. \\
 \hline
 FOUR PAWS International&Stray animals are starving in India, will you give them your much-needed support?&0.33 & w/o.\\
 \hline
 \multicolumn{3}{c}{\textbf{Category: Political actors}} \\
 \hline
 Arati Kreibich for Congress &Republicans are suppressing the vote through mass voter purges, polling place closures, and burdensome voter ID laws. Tell the Senate: restore the Voting Rights Act!&1&w.\\
 \hline
 Bernie Sanders&We are about to make history and I want you to be a part of it. Our campaign is trying to reach 1 million campaign donors faster than any campaign in American politics, and we are VERY close. Can you make a contribution right now to become one of our first million donors?&0.66&w.\\
 \hline
 Tina Smith&Meet Senator Tina Smith: a big fan of dogs, donuts, and Minnesotans.&0.66&w/o.\\
 \hline
 Judge Brian Hagedorn&Click here to hear how an adopted daughter changed the Hagedorn family!&0.33&w.\\
 \hline
\end{tabular}
}
\label{tab:exmp_adv_ctg_diff}
\vspace{-4mm}
\end{table*}

\subsection{Analysis of advertiser categories}

%\paragraph{Method to gather advertiser categories}
To characterize advertisers we analyze their category.  Advertisers need to create a Facebook Page and select from a pre-defined list a category for their page such as ``Software Company'' or ``Political Party''~\cite{adv_ctg_instr}. We use the advertiser's ids available in the dataset to collect their category using the Facebook Graph API. %The main category appears below the page name. To get advertiser's categories(any other data from Facebook) using a python code, you need to register as a developer on Facebook and then have an access token. After that, we used the Facebook library for python to collect categories from different advertisers. 
Some pages no longer exist, we were able to extract categories for 6476 ProPublica advertisers (82\%).
Figure~\ref{fig:all_categories_disclaimer_new} plots the breakdown of the corresponding advertisers categories for \spa, \pa, \pnpa, \opa and \npa. %There are \oana{9.342 -- wrong number} unique advertiser categories in the ProPublica dataset,
We group similar advertiser categories (grouping details can be found in our supplementary material at http://lig-membres.imag.fr/gogao/www21.html). % while the category ``Other'' contains all categories that we did not group. 

Figure~\ref{fig:all_categories_disclaimer_new} shows that most \spa come from political actors (58\% w. and 48\% w/o. disc.), but a significant fraction of ads also come from NGOs (14\% w. and 21\% w/o. disc.), communities (4\% w. and 4\% w/o. disc.), and advocacy groups (3\% w. and 4\% w/o. disc.). In the \pa group, a smaller fraction of ads come from political actors (24\% w. and 25\% w/o. disc.), much more from NGOs (36\% w. and 37\% w/o. disc.), and we also see more ads from advocacy groups (6\% w. and 6\% w/o. disc.), news media (4\% w. and 4\% w/o. disc.), and communities (6\% w. and 5\% w/o. disc.). In the \pnpa group, only (1\% w. and 1\% w/o. disc.) of ads come from political actors, the majority (52\% w. and 58\% w/o. disc.) from NGOs and charity organizations (11\% w. and 5\% w/o. disc.), some ads come from news media (5\% w. and 3\% w/o. disc.) and businesses (5\% w. and 3\% w/o. disc.). In the \opa group, we see a similar diversity in the advertisers labeling their ads as political. \textit{Many countries' specific electoral legislation only regulate (and impose restrictions on) ads from political actors. However, we see that there is a wide range of advertisers pushing political ads online and that volunteers do label ads from these advertisers as political; hence, prompting for updating legislation.}  

%\textit{These results show that there is a wide range of advertisers pushing political ads online and that, in many cases, volunteers agree on labeling ads non-traditional political advertisers as political (i.e., the \spa). This is relevant for country-specific electoral legislation that often only regulate ads from political actors~\cite{XX}.}

Facebook is explicitly exempting news organizations from labeling their ads as political even if they are about political issues~\cite{facebook-political-ads}; however, yet do seem to consider these ads as political. \textit{This raises the question of whether ads from news media should be treated as political ads.  On one side, political journalism is different from political propaganda; on the other side, news media has been used as a tool to manipulate users, and many unauthentic news aggregators are emerging with the purpose of promoting a political agenda~\cite{pink-slime}.}

Table~\ref{tab:exmp_adv_ctg_diff} presents examples of political ads from different categories of advertisers such as community, NGO, or business. For each ad, the table shows the fraction of political votes divided by all votes from volunteers and whether the ad was labeled as political by the advertiser itself. 
The table shows that there is a wide diversity of ads getting labeled as political. For instance, we can see an ad from the ice-cream company ``Ben and Jerry'' (a business) that is inciting citizens to vote, and an ad from the ``Democratic Attorneys General Association'' (an NGO) that is asking people who should be the V.P. of Joe Biden. Such ads have a clear association with elections.  
In the table, we also see many ads, such as the ones from the ``World Food Programme'' and the ``USA for UNHCR'' (Charities), that address social issues but do not seem to have any evident association to elections or legislation. \textit{The critical point to recognize is that ads labeled as political can have a very different level of ``politicalness'', going from straight advocacy messages addressing abortion issues to ads merely asking for a donation to end world hunger.}

\begin{figure*}[ht]
 \centering
 \includegraphics[scale=0.4]{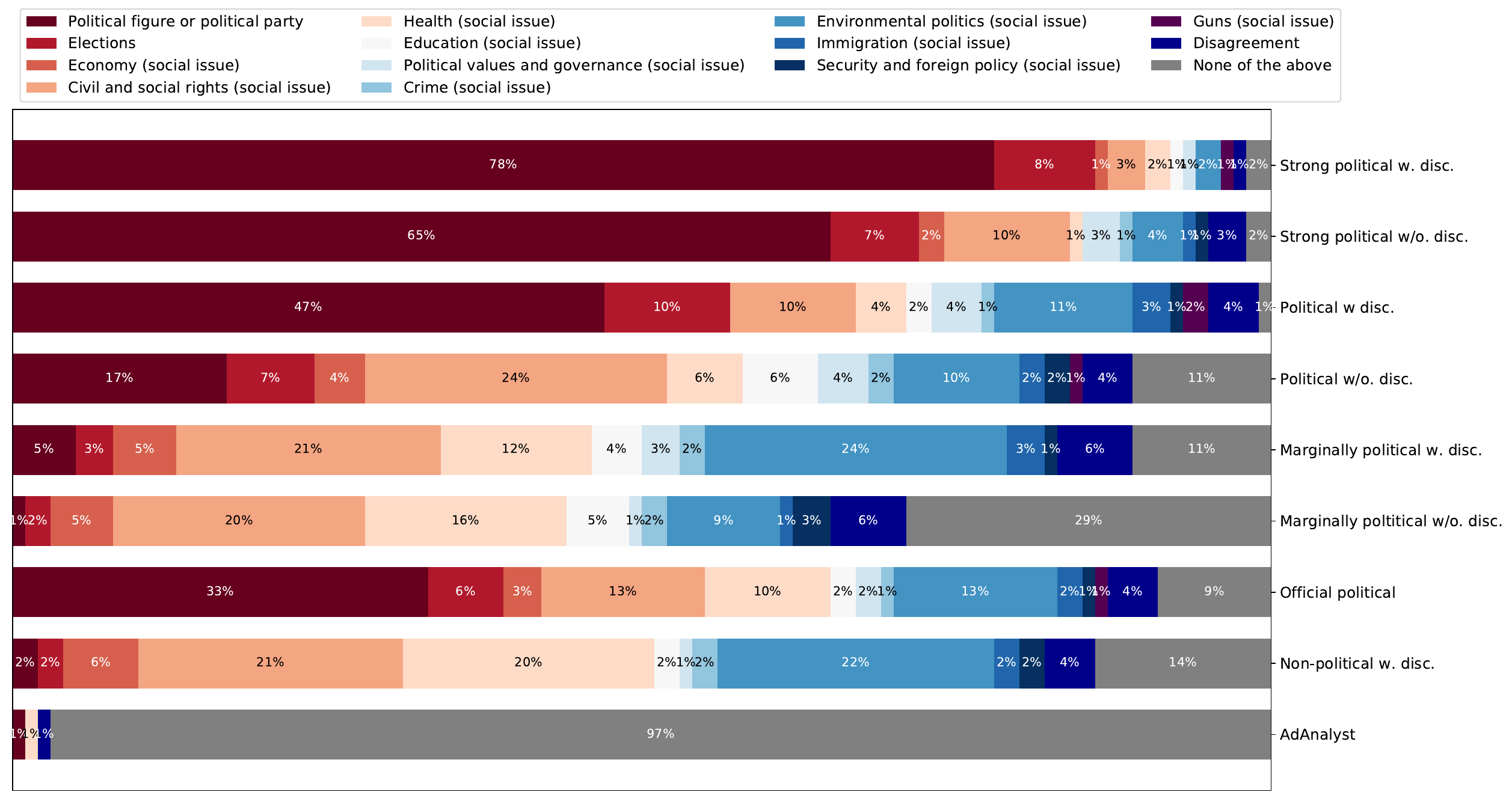}
 \vspace{-3mm}
 \caption{Breakdown of the political and social issues discussed in ads for the different groups of ads with and without disclaimer.}
 \label{fig:ad_message_type_disclaimer}
  \vspace{-4mm}
\end{figure*}

\subsection{Analysis of ad messages}
To gather grounded information about the topics of ads labeled as political, we took a random sample of 300 ads with a ``Paid for by'' disclaimer and 300 ads without "Paid for by" disclaimer from each \spa, \pa, \pnpa, 300 ads \npa, and 200 ads without disclaimer from AdAnalyst. %We picked both ads with and without disclaimer because volunteers might have been influenced by this visible tag when labeling ads as political. 
While we picked both ads with and without a disclaimer, we did not show the disclaimer in our surveys.
We set up a survey on Qualtrics~\cite{qlt} where for each ad, we ask respondents questions about the ad's message.
We hired workers through Prolific~\cite{prolific}, and we redirected them to fill out the survey on Qualtrics. Each worker had to label 20 random ads from the pool of 2300 ads, and each ad was labeled by three workers. We selected workers that are residing in the USA since all the ads used in the experiments targeted people who lived in or visited the USA. The median amount of time that workers spent on the survey was 12 minutes.

%\begin{figure}
% \centering
% \includegraphics[scale=0.1]{plots/wwf_ad.jpeg}
% \caption{Example of ad we asked workers to label.}
% \label{fig:sr_def}
% \vspace{-6mm}
%\end{figure}

Each survey had an instructions page, followed by 20 pages each containing one ad to label. For each ad, we asked the following questions:
%\vspace{-1mm}
%\begin{trivlist}
% \item \hspace{2mm} $\bullet$ ``Is this ad made by, on behalf of, or about a political actor? (such as a candidate for public office, a political figure, a political party or advocates for the outcome of an election to public office)'' 
%   \item \hspace{2mm} $\bullet$ ``Is this ad about elections? 
%(such as referendum or ballot initiative, including "go out and vote" or election campaigns)'' 
%   \item \hspace{2mm} $\bullet$ ``Does this ad refer to a social issue?  
%(such as civil and social rights, ...)'' 
%\end{trivlist}
%\vspace{-1mm}
(1) ``Is this ad made by, on behalf of, or about a political actor? (such as a candidate for public office, a political figure, a political party or advocates for the outcome of an election to public office)''; (2) ``Is this ad about elections? 
(such as referendum or ballot initiative, including "go out and vote" or election campaigns)''; and (3) ``Does this ad refer to a social issue? (such as civil and social rights, ...)''.   
Workers were allowed to answer yes to all the questions. If workers selected that the ad is about a social issue, we asked them which social issue: ``Which social issue is this ad talking about?'' Workers had to choose from the following list: civil and social rights, crime, economy, education, environmental politics, guns, health, immigration, political values and governance, security and foreign policy. We considered these social issues because they appear in the Facebook definition of political ads~\cite{soc_issue}. Workers were allowed to select multiple social issues if needed.  

If workers answered no for all three initial questions (the ad is not about a political figure, election, or social issue), they were asked to choose from a list "What topic describes best the ad". We took the list of 23 topics from the Interactive Advertising Bureau (IAB) categories~\cite{adv_ctg}. Note that we did not ask workers whether the ad is political or not; we just asked them questions about its message. 
Figure~\ref{fig:ad_message_type_disclaimer} shows the breakdown of the political or social issue discussed in an ad according to Prolific workers for different ad groups for ads with and without disclaimer. For each ad, we pick the ad topic chosen by the majority of workers or mark it as disagreement if no two workers chose the same ad topic or if two topics had an equal number of votes. We attributed all ads about both a political figure and a social issue or a political figure and election to the political figure group, and all ads about both an election and a social issue to the election group. For clarity, all ads for which the majority of workers chose a (non-political) IAB topic are marked as ``None of the above'' in Figure~\ref{fig:ad_message_type_disclaimer}.  
 
%\begin{figure*}
% \centering
% \includegraphics[height=0.15\textheight]{plots/newvers_ads_categories_4ctg.pdf}
% \vspace{-4mm}
% \caption{Types of ad message in different groups of ads. \oana{size of bars needs to be consistent across plots} \oana{we need to split in two: with disclaimer and without disclaimer}}
% \label{fig:ad_message_type_new}
% \vspace{-4mm}
%\end{figure*}

Figure~\ref{fig:ad_message_type_disclaimer} shows that all groups of ads contain most of the ad topics we consider. We see higher fractions of about a political figure or political party and ads about an election in the \spa (78\%+8\% w. and 65\%+7\% w/o. disc.) and higher fractions of social issues ads in the \pa (38\% w. and 61\% w/o. disc.) and \pnpa (75\% w. and 62\% w/o. disc.). In the \opa group, there is also a high fraction (48\%) of social issue ads.  The non-political AdAnalyst ads are shown as control. Indeed less than 2\% of these ads are labeled as being about a political figure, election or social issue. 
\textit{Social issue ads are only considered political by Facebook and TikTok, not by Google and Twitter. However, these results tell us that a large proportion of the ads volunteers and advertisers label as political are about social issues. Hence, it is crucial to consider social issue ads as political as well.}

Figure~\ref{fig:ad_message_type_disclaimer} shows that some ads (2\% w. and 2\% w/o. disc. of \spa and 1\% w. and 11\% w/o. disc. of \pa) were not labeled by workers as being about a social issue, a political figure or election. Since there is no expert ground truth, we cannot say whether labels from volunteers or labels from workers are better. 
 %Workers were also asked to choose a topic for the ad message even if they labeled the ad as non-political (the topics were selected from the IAB list~\cite{adv_ctg}).
Nevertheless, the (non-political) IAB topics that were mentioned the most by workers were society, health \& fitness, education and science. \textit{This raises questions on where to drawl the line between ads about civil and social right and ads about society; or ads about health as a social issue and ads about health \& fitness as a lifestyle.}  

One might decide that \pnpa should not be treated as political because only a minority of volunteers labeled them as political. 
%One might decide that all ads where the majority of votes are political should be considered political, while all ads where most votes are non-political should be considered non-political. Such an approach is, however, not adapted to political ads (e.g., \pnpa are not political).
Figure~\ref{fig:ad_message_type_disclaimer} shows that 5\%+3\% w. disc. and 1\%+2\% w/o. disc. of \pnpa do contain ads from a political figure or political party or elections. In addition, 21\% w. disc and 20\% w/o. disc. ads are about civil and social rights, and 24\% w. disc. and 9\% w/o. disc. are about environmental politics. The numbers look similar for \npa. 
Marginally political ads do contain a significant number of political ads as defined by the Facebook ToS. \textit{These results show that \pnpa should not be ignored because they might contain ads about social issue and ads where only a few people have the right background knowledge to detect them as political.} 
%Table~\ref{tab:exmp_pnp_ctg_elect} (in Appendix) shows a few examples of these ads. \oana{show some examples, takeaway. The point we want to make here is that in this category, there are ads about more local issues, and not everyone recognizes them}

\textbf{Takeaway:} Our results show that a large fraction of ads labeled as political are about social issues and do not mention a political actor or elections.  Hence, it is crucial to consider ads about social issues as political. 
Our results also show that a wide range of ads are getting labeled as ads about social issues.  Hence, since many legislative projects are considering to severely restrict micro-targeting~\cite{Jourova} or ban such ads altogether;  we need to decide whether we want ads (with no apparent link to elections and legislation) coming from charities or communities to be subject to the same restrictions as ads that advocate polarizing issues. Such restrictions could hurt a wide range of humanitarian civil organizations.

%While it is crucial to consider ads about social issue as political, we also need to decide whether we want ads (with no apparent link to elections and legislation) coming from charities or communities to be subject to the same restrictions as ads that advocate polarizing issues. While the current cost of labeling an ad as political is small (there are not many associated restrictions on Facebook--see Section~\ref{sec:agreement} for details); there are many legislative projects that are considering to severely restrict micro-targeting~\cite{Jourova} or ban such ads altogether. This could have a negative impact on a wide range of civil organizations.  

%% file: sources_disagreement.tex
%!TEX root = paper.tex

\section{Learning from disagreement}
\label{sec:sources_disagreement}

The previous section showed that a very diverse set of ads get labeled as political. This section analyzes the ads that lead to disagreement among volunteers and between volunteers and advertisers. 
This analysis is relevant for refining political ads' definition and improving the processes and instructions for labeling ads.

\begin{table*}
\caption{Ads underreported by advertisers or volunteers that are not about political figures, elections or social issues (according to workers).}
\vspace{-2mm}
\scriptsize{
% \begin{tabular}{|p{0.1\textwidth}|p{0.7\textwidth}|p{0.4cm}|p{0.4cm}|}
 \begin{tabular}{|p{0.15\textwidth}|p{0.65\textwidth}|p{0.07\textwidth}|p{0.02\textwidth}|p{0.015\textwidth}|}
 \hline
 Advertiser&Text & Workers' label & \fr & disc.\\  
 \hline
 
 \multicolumn{4}{c}{\textbf{Ads underreported by advertisers: \spa and \pa w/o. disclosure}} \\
  \hline
  
Citizens Against Lawsuit Abuse & Frivolous lawsuits are clogging our courts. Want to help tell trial lawyers enough is enough? Join Citizens Against Lawsuit Abuse (CALA) today!  & non-pol & 1 & w/o.\\
%\hline
%Indivisible Guide & If everyone reading this post signed up for a monthly donation to Indivisible -- even just the price of your morning coffee -- we?d have more than enough to support on-the-ground organizing. Can you commit to donating a little each month right now? It?ll make all the difference to our organization. & non-pol& 1 & w/o. \\
\hline
The Young Turks & Support independent investigative journalism while looking fly AF!  & non-pol& 1& w/o. \\
\hline
 Mikey Weinstein, MRFF & MRFF Op-ed: Anti-Theist Airman Memorialized by Air Force Unit with Image of Jesus  & non-pol &  0.75 & w/o. \\
%\hline
%Ketto & I have absolutely no savings and nothing left to sell. My son is fighting a tough battle with life and if I don?t arrange funds urgently, he?ll lose this fight. Only pure kindness can save him now,? the poor father pleads. The doctors have told them that his further treatment will costUSD 23,077.00, an amount far beyond their reach. They knew that no matter what they do, they?ll never be able to arrange the required sum. & non-pol &  0.66 & w/o. \\ 
\hline
Voices for Refugees & Torrential monsoon weather has hit Rohingya refugee camps in Bangladesh, destroying 273 family shelters already. Every donation helps us to reach those most vulnerable with emergency support and help to rebuild, reinforce and secure their shelters. & non-pol & 0.6 & w/o. \\
 \hline
  \multicolumn{4}{c}{\textbf{Ads underreported by volunteers: \npa and \pnpa w. disclosure}} \\
  \hline

Grist.org & Lettuce introduce you to the future of your arugula.  &  non-pol & 0 & w. \\
  \hline
  
Heifer International & Truth bee told, not everyone can get these 7 questions right. Test your bee smarts and unlock a 50 cent donation for Heifer  &  non-pol & 0 & w. \\
 % \hline
% & On average, how much honey does one bee produce in her lifetime? Take the quiz and find out!   &  non-pol & 0 & w. \\
  \hline
EveryLibrary  & Join hundreds of thousands of Americans who love libraries! &  non-pol & 0 & w. \\
  \hline
Mercy For Animals & Animals at factory farms suffer in unimaginable ways. They are cruelly confined, abused, neglected, and mutilated. Please support our work to stop this torment.   &  non-pol & 0 & w. \\
  \hline
The Christian Science Monitor & He need to address corruption in the Arab world is urgent. But if new initiatives are simply politically expedient – as many citizens suspect – they risk only fueling distrust and suspicion.   &  non-pol & 0 & w. \\
  \hline
 % Medi-Share  & With Medi-Share, your family has 24/7/365 access to telehealth doctors! Virtual care: any time, any place! &  non-pol &  & w. \\
 % \hline
 % Catholic Charities of Boston.  & Support a cause that does more for 165,000 people in Massachusetts. &  non-pol &  & w. \\
 % \hline
 % Feed the Children & By giving today, you can provide food and essentials to those in need. &  non-pol &  & w. \\
%  \hline
%Donate This Ramadan to Help Children Fight COVID-19 & Ramadan is bittersweet for the Syrian, Yemeni and Rohingya families spending another year...  &  non-pol &  & w. \\
 % \hline
 % CRTV  & If you want to be on the front lines of the fight for free speech and the open exchange of ideas, you need to get CRTV. You'll get the excerpt on social media, but if you want the full story, there's only one place for you.  &  non-pol &  & w. \\
%  \hline
 
 \end{tabular}
}

\label{tab:exmp_underreported}
\vspace{-5mm}
\end{table*}

\subsection{Volunteers vs. advertisers}
To understand why advertisers and volunteers disagree on ads being political, we examine separately ads that seem to be underreported by advertisers and ads that seem to be underreported by volunteers. 
 
\paragraph{Ads underreported by advertisers} These are the \spa and \pa without disclaimer. 
Table~\ref{fig:all_categories_disclaimer_new} shows that 4\% of the \spa and 7\% of the \pa are not labeled as political by advertisers. There are several possible (non-exhaustive) explanations: (1) advertisers do not comply with the ToS (e.g., they willingly do not label their ads as political to avoid scrutiny), i.e., volunteers are right; (2) advertisers underreport certain categories of political ads, i.e., advertisers and volunteers have different interpretations of which ads are political; and (3) volunteers misinterpret the ads' message, i.e., advertisers are right. %\footnote{This list is not exclusive, and there might be other explanations we did not consider.} 
 
Figure~\ref{fig:all_categories_disclaimer_new} presents the breakdown of advertiser categories and Figure~\ref{fig:ad_message_type_disclaimer} the breakdown of ad types corresponding to \spa, and \pa without disclaimer. A significant fraction of advertisers are political figures (48\% in \spa and 25\% in \pa), and a significant proportion of ads refer to a political figure or political party and elections (65\%+7\% for \spa and 17\%+7\% for \pa). 
 \textit{Hence, more than half of \spa and \pa without disclaimers are not compliant with Facebook's ToS. These results  confirm previous findings that advertisers sometime do not label their ads as political and the need for better enforcement mechanisms~\cite{silva2020facebook}.} 

A large fraction of ads without a disclaimer (23\% of \spa and 61\% of \pa) are about social issues. Recall that we excluded from this category ads labeled as social issues but mentioning a political figure or elections. 
Tables~\ref{tab:exmp_pnpa_civ}~and~\ref{tab:exmp_pnpa_env} (in Appendix) show some examples of ads about civil and social rights and environmental politics in \spa and \pa without disclaimer.  These ads are on topics such as climate change and healthcare, which are very politicized issues in the US, and give valid reasons to volunteers to label them as political.     

%Ads about social issues are more confusing than ads about political actors or elections; hence, advertisers might have underreported such ads without willingly trying to avoid detection.

To understand whether ads about some social issues are less disclosed by advertisers than others, for each ad topic, we compute the fraction of ads that do not have a disclaimer in the \spa and \pa groups. Ads about economy (0.15), civil and social rights (0.28), and security and foreign policy (0.27) have the lowest fraction of ads with a disclaimer. In contrast, ads about political figures (0.6), election (0.57), and environmental politics (0.49) have the highest fractions of ads with a disclaimer. \textit{This shows that advertisers are underreporting ads about social issues, especially if they are about economy or civil and social rights.}  

For 2\% of \spa, and 11\% of \pa w/o. disc. workers did not label them as being about a political figure, election, or social issue; which means that no one besides volunteers labeled them as political. Table~\ref{tab:exmp_underreported} shows a few examples of such ads. These ads seem to address some issues but are not clearly related to the social issues provided to workers.  This raises an interesting dilemma: if someone labels an ad as political (without being forced or by mistake), can they be wrong?

\paragraph{Ads underreported by volunteers} These are \npa and \pnpa with disclaimer. 
There are 1.3k \npa, and 5.6k \pnpa (74\%) labeled as political by advertisers.
There are various reasons why advertisers would label their ads as political while all/most volunteers labeled them as non-political: $(1)$ advertisers might be \textit{forced} to label ads as political (even if they are not) because of false positives in the enforcement mechanisms implemented by the ad platform; $(2)$ advertisers might think that disclaimers would bring more attention to their page; $(3)$ advertisers understand better why their ads should be political, and volunteers underreport such ads; etc. 
Figure~\ref{fig:ad_message_type_disclaimer} shows that a significant fraction (14\%) of \npa are labeled as not being related to a political figure, election or social issue by workers; meaning that no one besides advertisers are considering these ads as political. Table~\ref{tab:exmp_underreported} shows a few examples of such ads. Indeed, the majority of these ads do not seem to be political. \textit{Since substantial restrictions are envisioned for political ads, it is essential to know what enforcement mechanisms are put in place by ad platforms to understand what is the impact of false positives in their algorithms}. Non-political ads mislabeled as political is also problematic when building detection methods that use political ads labeled by advertisers to train models. Thus, it is important to look for poisoning attacks when building such models. 

Figure~\ref{fig:ad_message_type_disclaimer} shows that the majority of \npa and \pnpa without disclaimer are related to civil and social rights (21\% and 20\%), health (20\% and 16\%) and environmental politics (22\% and 9\%), while only a few refer to political actors (2\% and 1\%) or elections (2\% and 2\%). Figure~\ref{fig:all_categories_disclaimer_new} shows that these ads come mostly from NGOs (53\% and 58\%), news media (4\% and 3\%), businesses (4\% and 3\%), and charities (16\% and 5\%), while only a few (1\% and 1\%) come from political actors. 
\textit{Hence, it seems that volunteers underreport many ads about a social issue, especially about civil and social rights and health, and ads from advertisers such NGOs, and charities.}  
Table~\ref{tab:exmp_pnpa_civ}~and~\ref{tab:exmp_pnpa_env} (in Appendix) present examples of ads about civil and social rights and environmental politics in \npa and \pnpa.
We see that most of these ads are related to social issues, but volunteers might not consider them as political because there is no apparent association with elections or legislation. 

\textbf{Takeaway:} 
Two main factors contribute to disagreement between advertisers and volunteers: (1) advertisers mislabel ads as political or non-political (maybe to avoid scrutiny; maybe because they are forced to label their ads as political by enforcement mechanisms put in place by ad platforms); and (2) both advertisers and volunteers underreport ads about social issues. Part of the problem may be that the definition of ads about social issues is too broad which leads to different interpretations among people. This raises the question of whether \textit{all} ads related to social issues should be considered political, and if not, how should we filter social issue ads that are not political. For example, one possibility would be to consider as political only ads about social issues that could directly or indirectly impact elections or legislation or that address polarizing issues.
\subsection{Volunteers vs. volunteers}
\begin{figure*}[t]
  \centering
  \includegraphics[scale=0.5]{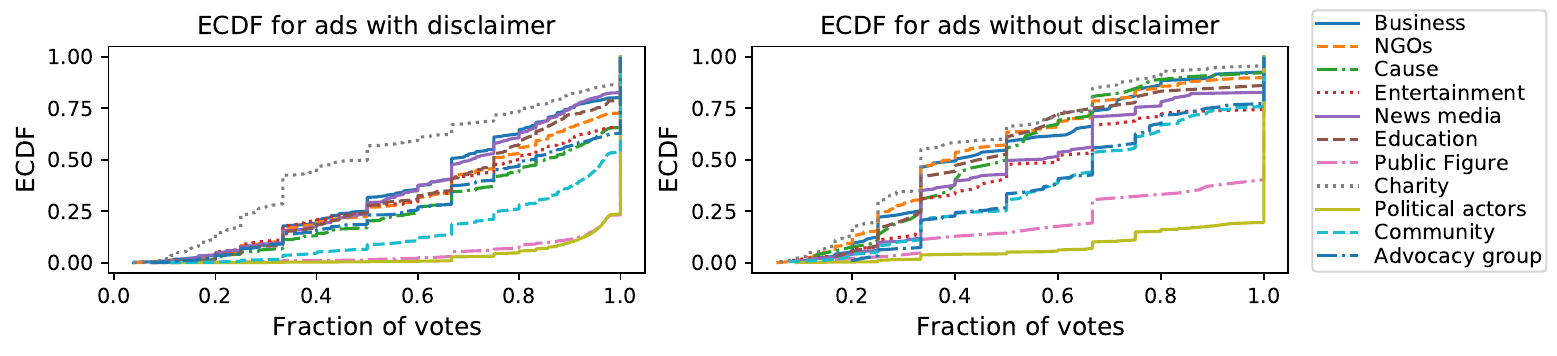}
      \vspace{-2mm}
  \caption{ECDF of the fraction of political votes for ads from different advertiser's categories in \spa, \pa, and \pnpa.}

  \label{fig:ECDF_fr_votes_adv_cat_all}
  \vspace{-2mm}
\end{figure*}

\begin{figure*}
  \centering
  \includegraphics[scale=0.50]{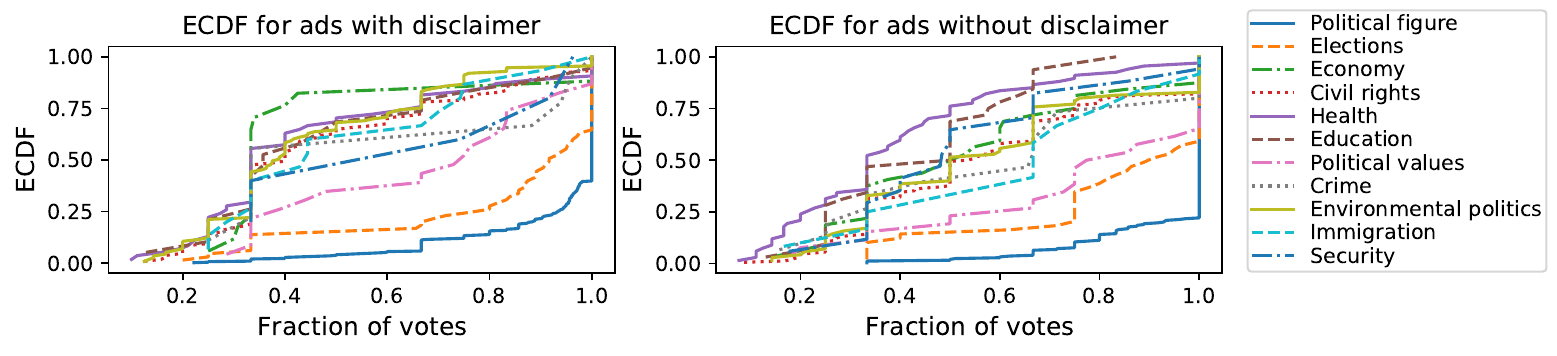}
      \vspace{-2mm}
  \caption{ECDF of the fraction of political votes for ads with different ad topic in \spa, \pa, and \pnpa.}

  \label{fig:ECDF_fr_votes_ads_label}
    \vspace{-4mm}
\end{figure*}

To investigate which ads lead to disagreement among volunteers, we check if there is more disagreement on ads coming from specific advertisers and ads about particular political or social issues. 

To see if ads from certain categories of advertisers lead to more disagreement, for each advertiser category, we group all corresponding ads (from \spa, \pa, and \pnpa). Figure~\ref{fig:ECDF_fr_votes_adv_cat_all} shows the ECDF of \fr for each group. Ads with a $\fr$ close to 0.5 have the highest level of disagreement (half of the volunteers label them as political and half as non-political). % \update{A vote is a strong signal for either considering an ad as political or non-political because the users themselves trigger the action of voting.} 
We split the analysis on ads with disclaimer and ads without a disclaimer since the disclaimer might have impacted how volunteers voted. We see in Figure~\ref{fig:ECDF_fr_votes_adv_cat_all} that the distributions shift to the right (more political votes) when ads have a ``Paid for by'' disclaimer. However, we cannot attribute this shift solely to the presence of disclaimers because ads with disclaimers might also have messages that are ``more political''. 
The plot shows that at least 10\% of ads in each advertiser category has $fr=1$, which means that at least \textit{some} volunteers are not bothered by the fact that the ad is coming from non-traditional political actors.
Figure~\ref{fig:ECDF_fr_votes_adv_cat_all} shows that ads coming from political actors and public figures achieve the highest agreement, 85\% have $fr=1$. Besides, ads from communities and advocacy groups tend to be seen as more political, while ads from charities as less political. Ads from other advertisers such as NGOs, causes, news media, education, and businesses are somewhere in between, leading to the highest level of disagreement. %This could be because they come from non-traditional political actors, or because the ads' message is not seen as political. 
To get definite proof if the advertiser category influences the decision (and not the message of the ad), we would need a conjoint analysis that tests the same ad message with different advertisers but our data does not permit such analysis. 
In any case, \textit{platforms and policymakers should clarify how much consideration should be given to the advertiser when labeling ads as political.}

To see if ads from certain ad topics lead to more disagreement, for each ad type, we group all corresponding ads (from the 1800 ads labeled by Prolific workers in \spa, \pa, and \pnpa). Figure~\ref{fig:ECDF_fr_votes_ads_label} shows the ECDF of \fr for each group (we only show groups for which we have more than 20 ads labeled).
%Figure~\ref{fig:ECDF_fr_votes_ads_label} presents the ECDF of $fr$ for ads with different ad message types for ads in \spa, \pa and \pnpa.  
We can see that the highest agreement is among ads that mention political figures and elections, while, as expected, the highest disagreement is on various social issue ads. 
We performed a pairwise Kolmogorov-Smirnov statistical tests between the distributions. Ads about elections and political figures are statistically different than the rest; but most of the social issue ads are not statistically different between them.
To see why for a particular social issues, some ads have higher \fr than others, 
Tables~\ref{tab:exmp_pnpa_civ}-\ref{tab:exmp_pnpa_env} (in Appendix) show examples of civil and social rights ads and environmental politics ads for different ad groups.We see that the ads address a wide range of topics (e.g., abortion, wildlife, hunger), they call for various actions (e.g., sign petitions, surveys, donations, call an elected representative) and try to provoke various sentiments (e.g., pride, anger, fear). Ads that address climate change and pollution are seen as more political, while ads about wildlife protection are seen as less political. Besides, ads that refer to problems in the U.S. (ad from NRDC) are seen as more political than ads that refer to problems in other countries (ad from Care2). While these are only anecdotal examples, they emphasize the complexity of deciding which ads are political.

\textit{Limitation:} There are other reasons for disagreement that we could not analyze with this dataset. For example, the background knowledge of volunteers might impact how they vote (the political nuance of an ad is only recognized by some) or the political ideology of volunteers impacts how they vote. %Our dataset does not have information about which volunteers voted for an ad, and we do not know how many ads a volunteer has voted. 
These questions are essential for recruiting moderators, and we leave them for future work. 

\textbf{Takeaway:} Ads from NGOs, causes, news media, education, and businesses and ads on social issues lead to the highest disagreement among volunteers. 
To distinguish better political from non-political ads, we would need policy recommendations that clarify the perimeter of social issue ads. This raise a multitude of questions such as: Should we treat ads about more politicized issues differently than ads about less politicized issues? Should we treat social issues depending on the country? Should we treat ads that call for precise actions differently than ads that just inform citizens? Should we define social issues at a smaller granularity (in both topics and locality) than currently? How should the system adapt to emerging social issues? 
%Should we treat ads addressing local social issues differently than national or word-wide social issues? 
How much weight should be given to the advertiser's identity (as opposed to just the ad content)? 

%% file: classifiers.tex
\section{Classification and disagreement}
\label{sec:classifier}
Traditional supervised classification algorithms create models from positive and negative examples that we feed in the training phase. The previous sections showed significant discrepancies between ads labeled as political by advertisers and ads labeled as political by volunteers. Hence, this raises the question of whether classifiers trained on one or the other would result in significantly different models. 
 %the question is  whether a classifier trained using positive labels from advertisers would result in a similar model with a classifier trained with positive labels from volunteers. 
Intuitively, if the training examples are biased, the models will be different, while if the training examples are representative of political ads in general, the resulting models will be similar.  This section investigates how discrepancies in positive labels from advertisers and volunteers impact the resulting classification models. 

For the evaluation we split the ProPublica dataset in two equal-size slices of 28k ads: $S_1$ and $S_2$. We use $S_1$ as the training and validation dataset and $S_2$ as the holdout/test dataset. We build four models using four different sets of positive labels but the same negative labels. As negative examples, we took 7.5k ads in English from AdAnalyst without the "Paid for by" disclaimer (see Section~\ref{sec:dataset}). 

\begin{trivlist}
%\vspace{-1mm}
 \item \hspace{2mm} \textbf{The $M_{op}$ model}: the positive labels are a random sample of 8000 \opa from $S_1$. $M_{op}$ is trained with positive examples from \textit{advertisers}. 
 \item \hspace{2mm} \textbf{The $M_{sp}$ model}: the positive labels are a random sample of 8000 \spa from $S_1$. $M_{sp}$ is trained with only positive examples where all volunteers agree they are political, $\fr=1$. 
 \item \hspace{2mm} \textbf{The $M_{mp}$ model}: the positive labels are a random sample of 4000 \pa and 4000 \spa from $S_1$. $M_{mp}$ is trained with positive examples where the majority of volunteers consider the ads as political, $\fr>0.5$. 
 \item \hspace{2mm} \textbf{The $M_{1p}$ model}: the positive labels are a random sample of 2600 \spa, 2600 of \pa and 2600 \pnpa from $S_1$. $M_{1p}$ takes as positive examples all ads where there exist at least one user that labeled it as political, $\fr>0$.
\end{trivlist}

To build the different models, we used Naive Bayes. While Naive Bayes is neither new nor sophisticated, it was shown by ~\citet{silva2020facebook} that it achieves high accuracy for detecting political ads and outperforms other methods. The classifiers only take as input the ad's text, and as pre-processing, we deleted all Html tags, stop words, and punctuation. We used Count Vectorizer for text embedding~\cite{pedregosa2011scikit}. 

We performed 10-fold cross-validation for each classifier over its specific training-validation dataset that contains 8000 positive and 7.5k negative examples. 
Table~\ref{tab: classifier} presents the average accuracy and true positive rate for a 1\% false positive rate for the four classifiers. For systems where the fraction of positive examples (political ads) is much smaller than the fraction of negative examples (non-political ads), it is essential to limit the rate of false positives (non-political ads labeled as political); hence, we are interested in true positive rates for a 1\% false positive rate.
The table shows that all classifiers achieve high accuracy of over 95\%, but only $M_{op}$, $M_{sp}$, and $M_{mp}$ achieve true positive rates of more than 90\%. The lower true positive rate of  $M_{1p}$ (85\%) is expected as it has a more challenging task because it is trained and tested with more debatable political ads.  
 
The main challenge in evaluating the classifiers is that we do not have a gold standard collection of political and non political ads. 
Table~\ref{tab: classifier} only tells us \textit{how good  the models are at identifying the same kind of political ads with the ones they were trained on, but not how good they are at identifying political ads in general}. Hence, we look next at how these models perform on detecting other kinds of political ads then those they were trained on.  

\begin{table}[]
 \caption{The average accuracy and true positive rate for a 1\% false positive rate for the four models. Each classifier is evaluated over its specific training-validation dataset.}  %Accuracy is computed on 10-fold cross validation over the specific training-testing datasets of each classifier.}
 \vspace{-2mm}
\small{
\begin{center}
 \begin{tabular}{|l|l|l|}
 \hline
   Classifier & Accuracy & TPR for 1\% FPR \\
  \hline
 \color{black}{$M_{op}$ model }& 96\% (+/-1\%) & \color{black}{92\% (+/-5\%)}\\
   \hline
 \color{black}{$M_{sp}$ model} & 97\% (+/-1\%) &96\% (+/-2\%)\\
   \hline
  \color{black}{$M_{mp}$ model}& \color{black}{95\% (+/-2\%)} & \color{black}{90\% (+/-4\%) }\\
   \hline
  \color{black}{$M_{1p}$ model }& \color{black}{95\% (+/- 3\%)} &\color{black}{85\% (+/-8\%)}\\
   \hline
 \end{tabular}
\end{center}

 \label{tab: classifier}
 }
 \vspace{-5mm}
\end{table}

We use the four models to make predictions for all ads in $S_2$. To predict that an ad is political, we took the threshold corresponding to a 1\% FRP for each classifier. Table~\ref{tab:classifier gp} shows how well the four models are at identifying \opa, \spa, \pa, \pnpa, and \npa in $S_2$. As negative examples, we used 1000 ads from \AdAnalyst  that do not have a disclaimer and were not used for training.  

Table~\ref{tab:classifier gp} shows that $M_{sp}$ has the lowest number of false positives, while $M_{op}$ has the largest number. 
For detecting \spa, all models detect more than 94\% of ads. For detecting \pa the $M_{op}$ and $M_{mp}$ perform the best (detecting over 94\% of ads). For \pnpa, $M_{1p}$ and $M_{mp}$, perform well (over 86\% detection), while $M_{1p}$ has a 84\% detection. For \npa, $M_{op}$ and $M_{mp}$ label more than 85\% as political.

The detection rates of $M_{op}$ and $M_{mp}$ are similar across different datasets, with $M_{mp}$ performing better especially on \pnpa and \npa. Hypothetically if the resulting classifiers would label as political the precise same ads, the input data is representative of political ads, and who is labeling  the training data (be it advertisers or volunteers) does not matter. 
To understand whether $M_{op}$ and $M_{mp}$ label the same ads as political, we computed the fraction of ads labeled by both models as political over all ads for different ad groups in $S_2$. The data shows that the two models have an overlap of 97\% in \spa, 94\% in \pa, 83\% in \pnpa, and 82\% in \npa. \textit{These results show that the overlap is relatively high, but discrepancies in the input data do transfer to discrepancies in the output data. Hence, we need to consider how biases in labeling are impacting classification results and whether this may lead to unfairness against certain advertisers.}

\begin{table}
\caption{The fraction of ads labeled as political by the four models in different groups of ads in $S_2$.}
 \vspace{-2mm}
\small{
 \begin{tabular}{|p{0.18\textwidth}|c|c|c|c|c|c|} 
 \hline
 & \# ads & $M_{op}$ & $M_{sp}$ & $M_{mp}$ & $M_{1p}$ \\ [0.5ex] 
 \hline\hline
 \textbf{AdAnalyst(non-political)} & 1000 & 1.3\% & 1.1\% &1.6\% &1\%\\
 \hline
 \hline
 \color{black}{\opa} & 25k & 95\% &90\% &97\%&91\%\\
\hline
 \color{black}{\spa} &13k& 97\% &97\% &98\%&94\%\\
\hline
\color{black}{\pa}& 8.7k&94\% &91\% & 97\% & 90\% \\
\hline
\color{black}{\pnpa} & 3.7k & 86\%&67\% & 89\% & 84\%\\
\hline
\npa & 0.6K & 85\%& 60\%&89\%&83\%\\
 \hline
\end{tabular}
}
\label{tab:classifier gp}
%\vspace{-4mm}
\end{table}

%% file: relatedworks.tex
%!TEX root = paper.tex

\section{Related Work}
%In this section, we provide an overview of studies related to political ads and their regulation in social media.
\noindent \textbf{Definitions of what is political:}
Several works studied what people think is political~\cite{whatispolmean,snapjudg}. \citet{whatispolmean} set up several experiments to identify what topics people consider political and showed that, similar to our study, it is complicated to defined what topics are political. The experiment included 33 different topics such as education, poverty, national parks and space exploration. The participants achieved a 95\% agreement that diet pills is not a political topic. However, the mean percentage of respondents who view a topic as political is 42\%. \citet{snapjudg} designed an implicit association test featuring the Supreme Court and Congress. The results showed that people perceived the Supreme Court as less implicitly political than Congress.
On a more theoretical side, \citet{polit_def} considered the question of the autonomy of politics. The author concluded that the current situation of politics is reflected in three different ways: outright extinction, autonomy or weakening, which leads to different ways of perceiving, identifying, and defining politics. \citet{whatispol} proposed that the concept of politics should help to clarify normative interests in politics, that the definition of politics should embrace everyday understandings of politics, and serve explanation. He suggested that politics can be define by two attributes: power and conflict. 

% Nowadays, science like of society-sociology-tends to absorb political science. The author concluded that the current situation of politics is reflected in three different ways: heteronomy, autonomy or dilution. Each way leads to different ways of perceiving, identifying, and defining politics. \citet{whatispol} proposed what concept of politics should do. First of all it should help to clarify normative interests in politics. Moreover, the definition of politics should embrace everyday understandings of politics. And finally it should define the domain of politics in ways that serve explanation. He suggested that politics can be define by two attributes: power and conflict. 

\vspace{1mm}
\noindent \textbf{Analysis of political conversations online:}
\citet{hersh2015hacking} described how political campaigns changed across time and concluded that social media has a large impact on peoples' decisions. \citet{maruyama2014hybrid} showed that Twitter activity could affect people's vote decision. The authors experimented during the 2012 U.S. Senate election in Hawaii. The results showed that people who actively participated in Twitter discussions changed their opinion about their candidate more often than people who did not use Twitter. \citet{kou2017one} analyzed the development of public discourses on social media. The authors showed that during the "Umbrella movement", conversations on Facebook (mostly used by Hong Kong citizens) empathized with protesters, while conversations on Weibo (mostly used by mainland China) empathized with the government.
% the differences in the discussion about Hong Kong’s "Umbrella movement" in two distinct social networks: Facebook (mostly used by Hong Kong citizens) and Weibo (mostly used by the mainland of China residence). The found out that discussions on Facebook were empathizing with the protesters while, discussions on Weibo were  empathizing with the government. 

\vspace{1mm}
\noindent \textbf{Political advertising:}
%Two recent studies attempted to detect political content and advertising. 
%\citet{oliveira2018politicians} proposed a CNN architecture to distinguish political tweets from non-political, where a tweet is considered political if a politician posts it and the content expresses subjects related to fundamental issues of state, politics, govern and justice. 
\citet{silva2020facebook} created a tool for collecting ads from Facebook and implemented several supervised classifiers to detect political ads during the 2018 election in Brazil. They detected a significant number of political ads that did not have the official ``Paid for by'' disclaimer.  % These two studies focus on detecting political content with high precision; however, they did not comment on the recall of the algorithms. 
\citet{10.1145/3287560.3287580} analysed ads send by the Russian interference in the 2016 US elections and found that ads were send to people less likely to report them.
\citet{edelson2020security} presented a clustering-based method to discover advertisers engaging in a potentially undeclared coordinated activity and proposed recommendations for improving the security of political advertising transparency. Furthermore, \citet{FBadtransp} summarized problems with the Facebook Ad Library, such as the lack of clear policies and data systematicity. Finally, \citet{ali2019ad} proved that Facebook's ad delivery algorithms effectively differentiate the price of reaching a user based on their inferred political alignment with the advertised content, inhibiting political campaigns' ability to reach voters with diverse political views. 
Our paper focuses on a more foundational question of what should be considered political advertising. %This building block is essential to build reliable classifiers to detect political ads and make political advertising more transparent. 

%\vspace{1mm}
%\noindent \textbf{Analysis of the Facebook ad ecosystem:}
Several other studies have pinpointed problems with the Facebook ad ecosystem without focusing on political advertising such as discrimination~\cite{speicher2018potential}, lack of transparency~\cite{andreou2019measuring,andreou2018investigating}, and security and privacy problems~\cite{8418598, korolova2010privacy}. 
%\citet{andreou2019measuring} analyzed who are the advertisers on Facebook and how are they using the system to target users. \citet{speicher2018potential} showed that advertisers could create discriminatory ads without using sensitive attributes. \citet{andreou2018investigating} showed problems with Facebook's ``why am I seeing this ad" explanations, which were often incomplete and vague. 

%% file: conclusions.tex
%!TEX root = paper.tex

\section{Conclusion}

Many agree that online advertising especially political adverting needs to urgently be regulated, but one missing key is how to reliably detect political ads. 
This papers attempts to dissect some of the complexity of labeling political ads. To our knowledge, this is the first study to show how ordinary people label ads as political, why they disagree and what are the implications for policymaking and enforcement algorithms. 

Our paper shows that volunteers seem to underreport ads from NGOs, and charities (that are considered political by advertisers) and advertisers seem to underreport social issue ads (that are considered as political by volunteers). 
%Our paper shows that volunteers seem to underreport ads from news media, NGOs, and business Moreover, we investigated how different training sets affect the classifier. The results show that the overlap between classifiers is the overlap is quite high, but on the negative side discrepancies in the input data do transfer to discrepancies in the output data.  Finally, the paper analyzed how different labelling instruction affected the labelling process. Providing the Facebook definition determined workers to see more political content in ads, however, the agreement does not seem to increase
While disagreement can be alleviated through better guidelines to a certain degree, many ads addressing societal and humanitarian issues are intrinsically hard to label. We believe that the community needs a gold standard collection for political ads and to better define the perimeter of social issue ads. %especially social issue ads to test workers and ad labeling processes. 
%Finally, even with the best definitions and guidelines, some ads will still be hard to label, hence, the questions is what would be some good algorithms that are able to handle well datasets with no clear ground truth. 
We hope our analysis can help policymakers and ad platforms to refine the definitions of political ads and their regulation.

%% file: appendix.tex
%!TEX root = paper.tex

\appendix
\section{Appendix}

\begin{table*}[h!]
\caption{Civil and social rights ads in different ad groups.}
\scriptsize{
% \begin{tabular}{|p{2cm}|p{0.7\textwidth}|p{0.4cm}|}
  \begin{tabular}{|p{0.15\textwidth}|p{0.75\textwidth}|p{0.015\textwidth}|p{0.015\textwidth}|}
 \hline
 Advertiser&Text & \fr & disc.\\  
 \hline
 \multicolumn{3}{c}{\textbf{Strong Political}} \\
 \hline
 AFSCME 3299 & Stand up for immigrant families.  Tell UC to cancel its contracts with ICE collaborators now! & 1 & w.\\
 \hline
    SEIU& Too many of us are still paid less for the same work. That’s why we need a union. & 1& w.\\
 \hline
  Fight Back& We're fighting for better healthcare and equal pay. & 1 & w/o \\
  \hline
  ACLU& Just a few weeks ago, the Supreme Court ruled that the First Amendment forbids religious hostility by the government. If only it applied that standard to the president and his Muslim ban.&1& w/o. \\
 \hline
  \multicolumn{3}{c}{\textbf{Political}} \\
 \hline
 CREDO Mobile&"We can transcend the darkness of this moment by joining the struggles of past and future freedom fighters. That is how, when we reach the end of our lives and look back on these heady moments, we will find peace in the knowledge that we did our best." – Ady Barkan.& 0.82 & w.\\
 \hline
 Granite State Progress Education Fund&Stop anti-abortion shame, stigma and hate from New Hampshire politicians. Sign the petition to support abortion access for all Granite Staters!&0.83&w.\\
 \hline
 Physicians for Human Rights&Doctors and nurses are standing up against human rights abuses across the world. Join our community and learn more about our work&0.75& w/o.\\
 \hline
International Rescue Committee&Women and girls in crisis zones face discrimination, violence, and a lack of equal opportunities. Learn how we're working to change that.& 0.66 & w/o.\\
 \hline
  \multicolumn{3}{c}{\textbf{Marginally Political}} \\
 \hline
Boston Rescue Mission &It’s tragic to be all alone and hungry. Your gift can bring hearty, nutritious meals to men and women who struggle with homelessness.& 0.2& w.\\
    \hline
  No Kid Hungry&Giving Tuesday is coming, and you can help end childhood hunger in America. Our partner, Citi, will match all donations up to \$100,000!&0.3&w.\\
    \hline
    International Rescue Committee &Yemen is facing the largest humanitarian crisis of our time: millions of children are at risk of starvation and a deadly cholera epidemic remains a serious threat. And it's about to get worse if we don’t step up our efforts now.& 0.2&w/o.\\
    \hline
    World Food Programme&Authorities in Yemen are blocking aid. Millions are suffering the consequences. Add your name today to keep aid flowing into Yemen&0.33&w/o.\\
    \hline
  \multicolumn{3}{c}{\textbf{Non-political}} \\   
    \hline
    Save the Children US&There’s still time to give during the 48 Hours of Giving! Your gift in support of  the Center for Girls will be matched 2x by an anonymous donor – but the match ends at midnight Saturday&0&w.\\
    \hline
    United States Holocaust Memorial Museum&It’s more important than ever that people understand the dangers of unchecked hatred. In this time of growing antisemitism at home and abroad, we all have a responsibility to keep the history of the Holocaust alive. Can we count on you?&0&w.\\
    \hline
    Covenant House International&TRIPLE your impact on precious young lives. Give now to help ensure that Covenant House keeps its pledge to welcome ALL homeless youth who come through our doors and love them unconditionally&0&w.\\
    \hline
    Nashville Rescue Mission&Water can be life-saving when summer’s heat is at its worst and there’s no escape. Helping is easy—and it won’t cost you a thing&0&w.\\
    \hline
\end{tabular}
}
%\vspace{-2mm}
\label{tab:exmp_pnpa_civ}
\end{table*}

\begin{table*}[h!]
\caption{Environmental politics ads in different ad groups.}
\scriptsize{
%  \begin{tabular}{|p{0.1\textwidth}|p{0.7\textwidth}|p{0.2cm}|p{0.2cm}|}
    \begin{tabular}{|p{0.15\textwidth}|p{0.75\textwidth}|p{0.015\textwidth}|p{0.015\textwidth}|}
 \hline
 Advertiser&Text & \fr & disc. \\  
 \hline
  \multicolumn{3}{c}{\textbf{Strong Political}} \\
  \hline
 National Parks Conservation Association& No organization has won more victories for the national parks over the past century than NPCA - but we can’t do it without you. Please donate to protect our nation’s magnificent public lands. & 1 &w.\\
 \hline
 Conservation Northwest & ACTION ALERT: Yesterday, the U.S. House of Representatives passed a budget bill that would block funding for grizzly bear restoration in the North Cascades. Use the links below to send your elected representatives a quick message to ensure Congress provides the funding bears need!& 1 & w.\\
 \hline
 Ocean Conservancy& Offshore oil spills can harm marine life, devastate ocean environments and risk the livelihoods of coastal communities.& 1 & w/o.\\
 \hline
 Coloradans for Responsible Energy Development& Colorado's first-in-the-nation oil and gas regulations work to protect our communities.&1 &w/o. \\
 \hline
  \multicolumn{3}{c}{\textbf{Political}} \\
  \hline
 Care2&Botswana is considering lifting the ban on hunting elephants. We must act NOW and convince Botswana to maintain their stance on protecting these endangered elephants from poachers!& 0.55 & w.\\
 \hline
American Bird Conservancy&The Endangered Species Act is under attack. Despite the fact that 99\% of species shielded by the Act — including Bald Eagles and California Condors — have avoided extinction, opponents in Congress are threatening to undermine this bedrock environmental law. Add your name to ABC's petition and tell the government to help protect endangered birds now&0.83 & w. \\
 \hline
NowThis & Women are equally impacted by climate change, and it’s critical that we have them equally involved in the solution& 0.66 & w/o.\\
 \hline
 NRDC& Plastics never break down. And that's becoming a real problem for those of us that depend on the Gulf of Mexico and Mississippi River& 0.8 & w/o.\\
 \hline

  \multicolumn{3}{c}{\textbf{Marginally Political}} \\
 \hline
Defenders of Wildlife&The support from our donors has helped us win many battles for wildlife, but there is always more to be done. Our love of animals is endless, so we are ready to fight tirelessly for imperiled wildlife that can’t speak for themselves. Support Defenders today and help us continue the fight for wildlife! &0.33 & w.\\
    \hline
    National Audubon Society Action Fund & Climate change threatens the birds we love. Sign up and we'll alert you to actions you can take to protect birds and the places we all need. & 0.17& w.\\
    \hline
  Potomac Conservancy&Trees are nature's Brita filters! For just \$33, we'll plant a native tree along the Potomac River to help filter out water pollution. Plant a tree today!& 0.36& w/o. \\
    \hline
    Climate Reality&Last year, 39 million people tuned in to 24 Hours of Reality to learn what climate change is doing to our planet and how we can solve it with the solutions in our hands today. Help us make 2018's show even bigger! & 0.33 & w/o.\\
\hline
 \multicolumn{3}{c}{\textbf{Non-political}} \\
    \hline
    National Park Foundation&Working together, you can help us have a powerful impact on our spectacular national parks. Your support right now will go to work immediately to protect the places that matter most for future generations.&0&w.\\
\hline
    The Nature Conservancy&The challenges facing our natural world are growing every day. Please, make a tax-deductible gift to give nature and wildlife a future.&0&w.\\
    \hline
    Oceana&Sea lions are drowning in mile-long "walls of death" off the California coast. Let them die... or help us save them&0&w.\\
\hline
IFAW&IFAW protects animals and the places they call home. With your help, we can continue to make a difference. Let's get to work.&0&w.\\
\hline
\end{tabular}
}
\label{tab:exmp_pnpa_env}
\vspace{-2mm}
\end{table*}